\documentclass[pre,twocolumn,aps,floats,showpacs,superscriptaddress]{revtex4}

\usepackage{amsmath}
\usepackage{amssymb}
\usepackage{latexsym}
\usepackage{bm}
   
\usepackage{graphicx}
\usepackage{epsfig}
\usepackage{epic}
\usepackage{eepic}

\newcommand{\dd}{\delta}

\begin{document}
\bibliographystyle{prsty}
\author{Balazs Kozma} 
\affiliation{LPT, CNRS, UMR 8627, and 
Univ Paris-Sud, Orsay, F-91405  (France)}
\author{Alain Barrat} 
\affiliation{LPT, CNRS, UMR 8627, and 
Univ Paris-Sud, Orsay, F-91405  (France)}
\affiliation{Complex Networks Lagrange Laboratory, ISI Foundation, 
Turin, Italy}

\title{Consensus formation on adaptive networks}
\widetext

\begin{abstract}
 
  The structure of a network can significantly influence the properties of the
  dynamical processes which take place on them. While many studies have been
  devoted to this influence, much less attention has been devoted to the
  interplay and feedback mechanisms between dynamical processes and network
  topology on adaptive networks. Adaptive rewiring of links can happen in real
  life systems such as acquaintance networks where people are more likely to
  maintain a social connection if their views and values are similar. In our
  study, we consider different variants of a model for consensus formation.
  Our investigations reveal that the adaptation of the network topology
  fosters cluster formation by enhancing communication between agents of
  similar opinion, though it also promotes the division of these clusters. The
  temporal behavior is also strongly affected by adaptivity: while, on static
  networks, it is influenced by percolation properties, on adaptive networks,
  both the early and late time evolution of the system are determined by the
  rewiring process. The investigation of a variant of the model reveals that
  the scenarios of transitions between consensus and polarized states are more
  robust on adaptive networks.

\end{abstract}
\date{\today}

\pacs{89.75.-k, -87.23.Ge, 05.40.-a}

%89.75.-k       Complex systems
%87.23.Ge       Dynamics of social systems
%05.40.-a Fluctuation phenomena, random processes, noise, and Brownian motion
\maketitle

\section{Introduction, motivation}

The recent past has witnessed an important development of the
activities of statistical physicists in the area of social sciences,
motivated by the fact that statistical physics is the natural field to
study how global complex properties can emerge from purely local
rules. Statistical physics models and tools have therefore been
applied to the understanding of
issues related to the characterization of the collective social behavior 
of individuals, such as culture dissemination, the spreading of linguistic 
conventions, and the dynamics of opinion formation~\cite{sociophysics}.

The statistical physics approach tries to grasp the essential
features of emerging social behaviors, and considers therefore simple rules of
opinion formation in which agents update their internal state, or opinion,
through an interaction with their neighbors. According to the ``herding
behavior'' described in sociology~\cite{Follmer:1974,Chamley:2003}, such
interaction typically consists of agents following local
majority~\cite{Glauber:1963,Galam:1982,Krapivsky:2003} or imitating a
neighbor~\cite{Krapivsky:1992}. Starting from random initial conditions, and
without any global supervision, the system self-organizes through an ordering
process possibly leading to the emergence of a global {\em consensus}, in
which all agents share the same opinion. Alternatively, the system can reach a
state of {\em polarization}, in which a finite number of groups with different
opinions survive, or of {\em fragmentation}, with a final number of opinions
scaling with the system size.

In certain models, opinions are represented very schematically by a discrete
variable which can take two values ($0$ or $1$), similarly to Ising spins;
this is the case in the Voter model \cite{Krapivsky:1992}, for which, at each
timestep, an agent is chosen at random and adopts the opinion of one of its
neighbors. Some additional realism on the modeling of opinions is put forward
in Axelrod's model \cite{Axelrod:1997}, where opinions or culture are
represented by a vector of cultural traits. Features such as memory may also
be introduced, with interesting new emerging behaviors
\cite{Steels:1996,Baronchelli:2006,Baronchelli:2006b,Dallasta:2006c,Castello:2006}.
Another refinement with respect to the use of binary opinions is introduced in
the Deffuant model \cite{Deffuant:2000} where opinions are continuous
variables (see also
\cite{BenNaim:2003,StaufferOrtmanns2003,Amblard:2004,BenNaim:2005}).  The
latter models also introduce the notion of bounded confidence: an agent will
interact with another agent {\em only if their opinions are close enough}. The
bounded confidence is described by a tolerance parameter, and the system can
evolve towards different states of polarization depending on the value of this
parameter.

As a first natural step, many studies of such simple models have considered
that each agent was allowed to interact with all the others. This
mean-field-like scenario can indeed be realistic when dealing with a small
number of agents. Moreover, the case of agents embedded into low-dimensional
lattices has as well been a topic of interest. Recently however, the growing
field of complex networks
\cite{Albert:2002,Dorogovtsev:2002a,Dorogovtsev:2003a,Newman:2003b,pastorsatorras2004}
has allowed to obtain a better knowledge of social
networks~\cite{Granovetter:1973,Wasserman:1994}, and in particular to show
that the typical topology of the networks on which social agents interact is
not regular. Various studies have therefore considered the evolution of the
aforementioned models when agents are embedded on more realistic networks,
and studied the influence of various complex topologies on the corresponding
dynamical behavior (see for example
\cite{Castellano:2005,Sood:2005,Suchecki:2005a,Suchecki:2005b}).

Up to now, few studies have however considered the fact that many networks
have a dynamical nature, and that their evolution occurs 
on a timescale which may
have an impact on the dynamical processes occurring between the nodes. Such
considerations are particularly relevant for social network which continuously
evolve, a priori on various timescales (both fast and slow). Moreover, the
evolution of the topology and the dynamical processes can drive each other
with complex feedback effects. The topology may indeed have an
impact on the evolution
of the agents' states, which in its turn determines how the topology can be
modified
\cite{Zimmermann:2004,Ehrhardt:2006,Gil:2006,Centola:2006,Holme:2006,Caldarelli:2006,Stauffer:2006}:
the network becomes adaptive.

In this paper, we therefore investigate how the coevolution of an adaptive
network of interacting agents and of the agents' opinions influence
each other, and how the final state of the system depends on this
coevolution. We focus on the Deffuant model for which a large number
of opinions can coexist (and not only 2 as in the Voter
Model). Moreover, and in contrast with most other studies of evolving
networks, the rate of evolution of the network's topology is tunable
and represents one of the parameters of the model. We focus on simple
evolution rules that do not require prior knowledge of the state of
agents to which new links are established. We investigate the role
of the various parameters such as the tolerance of agents and the
rate of topology evolution. We show that the possibility of the interaction
network to adapt to the changes in the opinion of the agents
has important consequences on the evolution mechanisms and on the
structure of the system's final state.

The organization of the paper is as follows: In Section \ref{sec:def}, the
Deffuant model and the investigated quantities are defined. In order to
have a full description of the model, we start our study with the
opinion formation of static networks in Section \ref{sec:SN}. Then, the
case of adaptive networks is considered in Section \ref{sec:AN} in
comparison with the static case. In this section, we study the
effects of the rewiring on the final state of the consensus formation.
Next, in Section \ref{sec:Temp}, the investigation of the temporal behavior
of the system gives us a deeper understanding of the processes taking
place of adaptive networks. Finally, in Section \ref{sec:variant}, a
variant of the original Deffuant model is considered on static and adaptive
networks.  

\section{Definition of the model and quantities of interest}
\label{sec:def}

The model we consider is based on the Deffuant model for interacting agents
\cite{Deffuant:2000}. In this model, $N$ agents ($i=1,\cdots,N$) are endowed
with a continuous opinion $o$ which can vary between $0$ and $1$ and is
initially random.  Two agents, $i$ and $j$ can a priori communicate with each
other if they are connected by a link, i.e. if they are neighbors. At each
time step $t$, two neighboring agents are selected, and they communicate if
their opinions are close enough, i.e., if $|o(i,t)-o(j,t)|\le d$, where $d$
defines the tolerance range or threshold. In this case, the (local)
communication tends to bring the opinions even closer, according to the rule
\begin{eqnarray}
\nonumber
   o(i,t+1)=o(i,t)+\mu(o(j,t)-o(i,t)) \\
   o(j,t+1)=o(j,t)-\mu(o(j,t)-o(i,t))
\label{deffuant}
\end{eqnarray}
where $\mu \in [0,1/2]$ is a convergence parameter. For the sake of simplicity
we will consider the case of $\mu=1/2$ which corresponds to $i$ and $j$
adopting the same intermediate opinion after communication
\cite{BenNaim:2003}. The role of the tolerance threshold has been
characterized in the mean-field topology where all agents are neighbors of
each other. For large tolerance values, agents can easily communicate and
converge to a global consensus. On the contrary, small values of $d$ naturally
lead to the final coexistence of several remaining opinions.

In the present study, we consider more realistically that agents have a limited
number of neighbors. The initial interaction network structure is taken as an
uncorrelated random graph in which agents have $\bar{k}$ 
acquaintances on average, i.e. the initial network corresponds to 
an Erd\H{o}s-R\'{e}nyi network with average degree $\bar{k}$.

In the next section, we will study for reference the case of a static
interaction network. This framework considers that the topology of the agents'
interaction does not evolve, or evolves at a rate which is infinitely slow
with respect to the communication between agents. It is however also
interesting to consider the fact that social interactions may evolve on the
same time scale as agents' opinions, and possibly in a way depending on these
opinions.  Agents indeed may break, keep or establish connections according to
how much frustration or reward they get from the corresponding 
relationship. The network along which communication and possible convergence
of opinions occur becomes then time-dependent. Many possibilities can be
thought of for modeling this time evolution: links may for example decay at
constant rate, independently from the agents' opinions \cite{Ehrhardt:2006}.
Within the framework of opinion dynamics with bounded confidence however, it
seems natural to consider that only connected agents having opinions which
differs more than the tolerance range may decide to terminate the
relationship. In order to keep the average number of interactions constant, a
new link is then introduced between one of the agents having lost a connection
and another agent. In our model, this new link is established at random
\footnote{Other models consider that the new connection is established towards
  an agent with a similar opinion \cite{Holme:2006}; since this requires an a
  priori knowledge of the new agent's opinion, and in fact therefore of the
  whole system, we stick to a simpler hypothesis of a randomly established new
  connection.}. Naturally, this new link can break again if the
newly connected agents have too far apart opinions. The rewiring process
thus occurs as a random search for agents with close enough opinions.

The model therefore considers two coexisting dynamical processes: local
opinion convergence for agents whose opinions are within the tolerance range,
and rewiring process for agents whose opinions differ more. The relative
frequencies of these two processes is quantified by the parameter $w \in
[0,1]$.  The precise rules of evolution are therefore summarized as follows.
At each time step $t$, a node $i$ and one of its neighbors $j$ are chosen at
random. With probability $w$, an attempt to break the connection between $i$
and $j$ is made: if $|o(i,t)-o(j,t)| > d$, a new node $k$ is chosen at random
and the link $(i,j)$ is rewired to $(i,k)$ \footnote{One can also think of a
  different rewiring rule in which $(i,j)$ is rewired to $(k,j)$. While the
  global qualitative picture does not change, the influence of this
  alternative rule deserves further investigation \cite{AGandBK}.}. With
probability $1-w$ on the other hand, the opinions evolve according to
(\ref{deffuant}) if they are within the tolerance range.  If $w>0$, the
dynamics stops when no link connects nodes with different opinions. This can
correspond either to a single connected network in which all
agents share the same opinion, or to several clusters representing
different opinions. For $w=0$ on the other hand, the dynamics stops
when neighboring agents either share the same opinion or differ of
more than the tolerance $d$.

Using a semi-formal algorithmic description let us rewrite the steps of
 the simulation:
\begin{enumerate}
\item Choose a node randomly, node $i$.
\item Pick one of its neighbors, node $j$.
\item Generate a random number, $r$, between $0$ and $1$.
\item \textbf{if} $(r>w)$ 
 \begin{description}
    \item \textbf{then} \textbf{if} $(|o(i,t)-o(j,t)|<d)$ 
          \begin{description}
            \item \textbf{then} opinion convergence for $i$ and $j$:
               \begin{description}
                 \item $o(i,t+1)=o(i,t)+\mu(o(j,t)-o(i,t))$ 
                 \item $o(j,t+1)=o(j,t)-\mu(o(j,t)-o(i,t))$ 
               \end{description}
          \end{description}
   \item \textbf{fi}
   \item \textbf{else} \textbf{if} $(|o(i,t)-o(j,t)|>d)$ 
          \begin{description}
            \item \textbf{then} update the link between $i$ and $j$:
               \begin{enumerate}
                 \item Choose a random node, node $k$, which is neither $i$
                 nor $i$'s neighbor.
                 \item Break the link between $i$ and $j$ then connect
                 $i$ with $k$.
               \end{enumerate}
          \end{description}
    \item \textbf{fi}
  \end{description}
  \textbf{fi}
\item Start again from Step 1.
\end{enumerate}

The evolution of the system and its final state can be characterized by the
investigation of the {\em opinion clusters} of agents.  In the final
state, such clusters are made of agents sharing the same opinion. During the
dynamical evolution however, many agents have close but not identical
opinions, so that we generalize the concept of opinion clusters in the
following way: two agents are considered as members of the same opinion
cluster if there is a path of agents in between them on the interaction
network where each consecutive agent in the path has an opinion within the
tolerance value of the previous agent. This corresponds to the idea that there
is a channel of communication in between them to share ideas. The notion of
opinion clusters gives a natural way to keep track of the structure of the
system over the whole dynamical process.

In the case of evolving (adaptive) networks, we also keep track of the 
{\em topological clusters} which correspond simply to the various
connected components of the network. In the final state, the topological
and opinion cluster naturally coincide, while for static networks, 
a unique connected cluster of agents can host several opinion clusters.

%%%%%%%%%%%%%%%%%%%%%%%%%%%%
\begin{figure}[thb]
\centering
\rotatebox{0}{
\includegraphics[width=.5\textwidth]{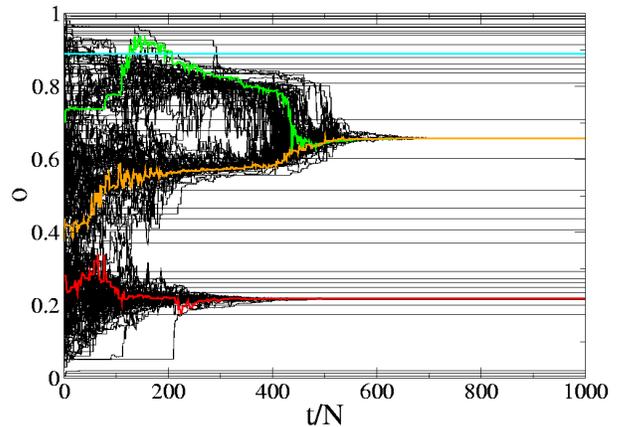}
}
\vspace*{-0.0cm}
\caption{ 
\label{fig:path_static}
(Color online) 
Evolution of the opinions of $25\%$ of the population, denoted by 
lines, for a system of $10^3$ agents with tolerance $d=0.15$ and 
average degree $\bar{k}=5$, on a static network for a single run. The
evolution of the opinion of a few individuals is highlighted with color.  }
\end{figure}
\begin{figure}[thb]
\centering
\rotatebox{0}{
\includegraphics[width=.5\textwidth]{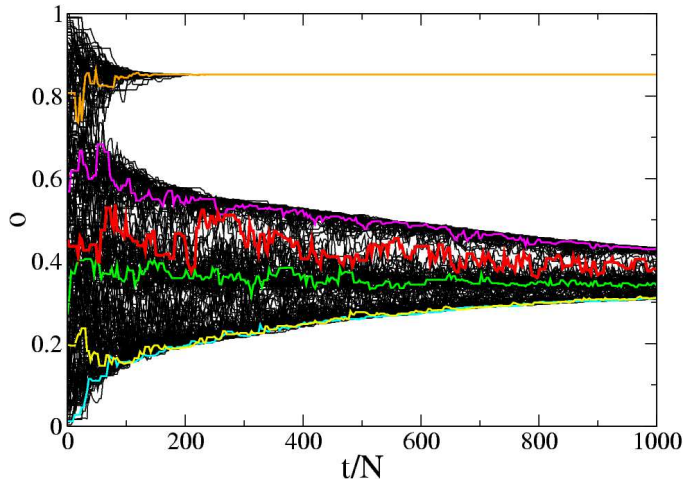}
}
\vspace*{-0.0cm}
\caption{ 
\label{fig:path_adaptive}
(Color online)
Same plot as for Fig. \ref{fig:path_static} for an adaptive network when
the rate of update is $w=0.7$ ($N=10^3$, $\bar{k}=5$, and $d=0.15$).
}
\end{figure}

%%%%%%%%%%%%%%%%%%%%%%%%%%%
Before turning to a detailed analysis of the model, we illustrate in
Fig.s \ref{fig:path_static} and \ref{fig:path_adaptive} the different
behaviors observed for static and adaptive networks. The figures show
the evolution of the opinions of $250$ out of $N=1000$ agents as a function of
time, in each case for one single realization of the dynamics with
$d=0.15$. The opinions are initially randomly distributed on the
interval $[0,1]$. When the interaction network is static, local
convergence processes take place and lead to a large number of opinion
clusters in the final state, with
few macroscopic size opinion clusters and many small
size groups: agents with similar opinions may be
distant on the network and not be able to communicate. This is in
contrast with the mean-field case in which all agents are linked together
so that the final opinion clusters are less numerous and more separated
in the opinion space. Figure \ref{fig:path_adaptive}, which corresponds
to an adaptive network with $w=0.7$, is strikingly in contrast with
the static case: no small groups are observed. We will investigate these
differences in more details in the next sections. 

%%%%%%%%%%%%%%%%%%%%%%%%%%%%
%%%%%%%%%%%%%%%%%%%%%%%%%%%

In particular, the whole cluster-size distribution gives a complete
description of the system. Interesting summaries are given by the {\em
number of clusters}, the {\em size of the largest} and {\em
second-largest opinion-cluster} which will tell us about the behavior
of the clusters with macroscopic size (because of the possible
large number of small clusters, the average size may be biased towards
small values and is therefore of less interest). 

\section{Consensus formation on static networks}
\label{sec:SN}

\begin{figure}[thb]
\vspace*{0.65cm}
\centering
\includegraphics[width=.45\textwidth]{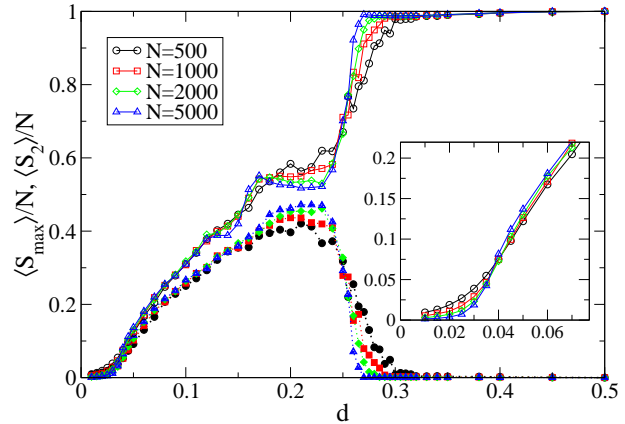}
\vspace*{-0.0cm}
\caption{ 
\label{fig:FSS_static}(Color online)
Size of the largest (empty symbols) and second-largest (filled
symbols) clusters in the final state, as a function of the tolerance
value on a static Erd\H{o}s-R\'enyi network with average degree
$\bar{k}=10$ for different system sizes. The color coding is the same
for the first and second-largest cluster with respect to the system
size. Inset: Same plot for the largest cluster zoomed into small
tolerance values near the polarized-fragmented transition. }
\end{figure}

Let us first consider for reference the Deffuant model on a static
Erd\H{o}s-R\'enyi network. Figure \ref{fig:FSS_static} displays the average
relative size of the largest ($\langle S_{max}\rangle/N$) and second-largest
opinion clusters ($\langle S_2\rangle/N$)) in the final state, as a function
of the tolerance parameter $d$, for various system sizes. Simulations are
averaged over 100 different networks. Three different phases can be readily
identified. At large tolerance values $d > d_1$, the system is in a {\em
  consensus} state, with a single macroscopic-size cluster present in the
final state. A jump of $\langle S_{max}\rangle/N$ from a value close to $1$ to
a value close to $1/2$ is observed around $d_1\approx0.256$, together with the
appearance of a macroscopic second largest cluster. This jump becomes sharper
and sharper when the system size increases, hinting at a first-order phase
transition in the thermodynamic limit.

%%%%%%%%%%%%%%%%%%%%%%%%%%%%
%%%%%%%%%%%%%%%%%%%%%%%%%%%

\begin{figure}[thb]
\vspace*{0.5cm}
\centering
\includegraphics[width=.45\textwidth]{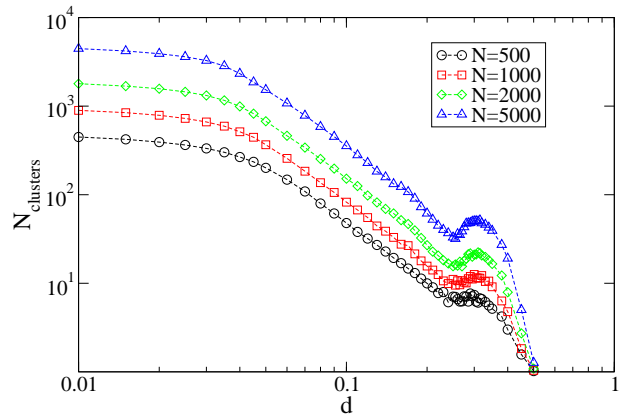}
\vspace*{-0.0cm}
\caption{ 
\label{fig:FSS_no_clust_static}(Color online)
Number of clusters in the final state
as a function of the tolerance of the agents for
different system sizes on a static Erd\H{o}s-R\'enyi
network with average degree $\bar{k}=10$.
}
\end{figure}

The evolution of the number of opinion clusters in the final state,
$N_{clust}$, sheds more light on the system's behavior, as shown in Fig.
\ref{fig:FSS_no_clust_static}. For $d > 0.5$, only one final cluster
containing all nodes is obtained. As $d$ decreases, $\langle S_{max}\rangle/N$
decreases very slowly while $\langle S_2\rangle/N$ remains close to $0$ and
$N_{clusters}$ increases: a large number of small clusters of finite size
appear. When $d$ approaches and crosses the transition point, an interesting
non-monotonic behavior is observed: $N_{clusters}$ decreases as the tolerance
of the agents decreases towards $d_1$. This corresponds to the appearance of a
second largest cluster of large size. This second large cluster contains
agents with opinion $o_2$ different from the largest cluster's opinion $o_1$.
The ``global'' tolerance range of these two large clusters is therefore wider
than if only one large cluster of agents with the same opinion is present: it
goes from $[o_{1}-d,o_{1}+d]$ to the union $[o_{1}-d,o_{1}+d] \cup
[o_{2}-d,o_{2}+d]$ and therefore allows to communicate with more agents and
obtain less small (finite size) clusters in the final state. The larger this
second cluster is, the less the finite-size clusters remain isolated, therefore
$N_{clusters}$ decreases.

%%%%%%%%%%%%%%%%%%%%%%%%%%%%
%%%%%%%%%%%%%%%%%%%%%%%%%%%

\begin{figure}[thb]
\vspace*{0.5cm}
\centering
\rotatebox{0}{
\includegraphics[width=.45\textwidth]{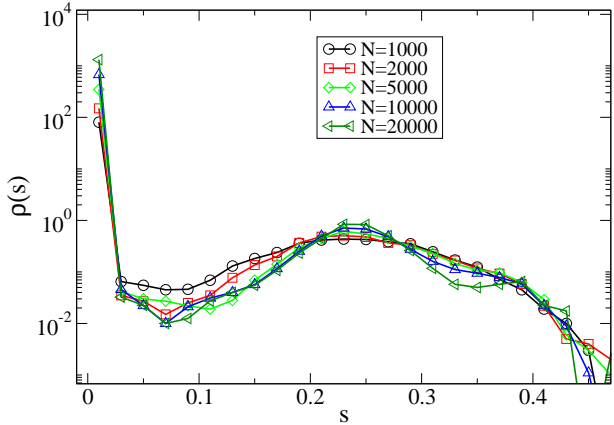}
}
\vspace*{-0.0cm}
\caption{ 
\label{fig:Pcs_static}(Color online)
Cluster size distribution
$\rho(s)$ for $d=0.1$ and $\bar{k}=10$ on a static Erd\H{o}s-R\'enyi
network with average degree $\bar{k}=10$.
}
\end{figure}

For $d < d_1$, an apparently polarized state
is entered, with a first and second-largest clusters of similar
extensive sizes. The investigation of the number of clusters 
(Fig. \ref{fig:FSS_no_clust_static}) however shows that the system is in
a {\em ``false''-polarized} state, in which the number of clusters increases
with the system size. This state therefore consists of a coexistence of
macroscopic opinion clusters with an extensive number of finite size
clusters. As $d$ decreases, the decrease of the sizes of the largest and
second-largest clusters is thus due to two reasons: the appearance of more and
more macroscopic-size clusters, as it is also the case in mean-field, and
the proliferation of finite size ``microscopic'' clusters. This
last point is made more explicit by the investigation of the whole
cluster-size distribution displayed in Fig. \ref{fig:Pcs_static}
for $d=0.1$. The figure shows that the distribution
of normalized sizes $s=S/N$ is composed of two parts
\begin{equation}
\rho_N^0 (s) \approx \dd(s)f^0(N) + Q^0(N,s),
\end{equation}
where $f^0(N)\propto N$ gives the number of isolated small clusters
and $Q^0$ is a regular part describing clusters of macroscopic size.

%%%%%%%%%%%%%%%%%%%%%%%%%%%%
%%%%%%%%%%%%%%%%%%%%%%%%%%%%

\begin{figure}[t]
\vspace*{0.5cm}
\centering
\includegraphics[width=.45\textwidth]{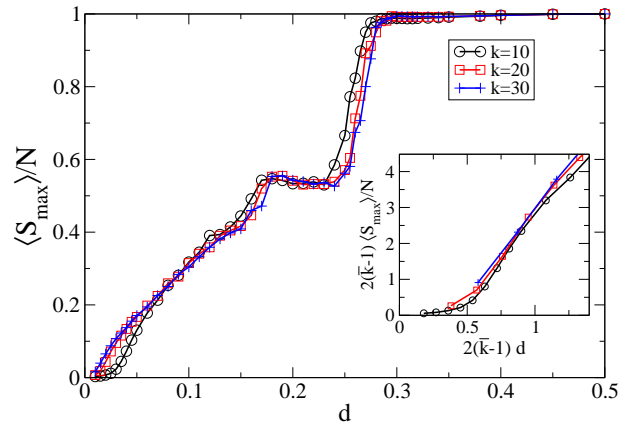}
\vspace*{-0.0cm}
\caption{ 
\label{fig:diff_ks_static}(Color online)
Size of the largest cluster as a function of the tolerance value on
static Erd\H{o}s-R\'enyi networks 
($w=0$) for different average connectivities. Inset: same plot, 
rescaled by the average connectivity and zoomed into the
fragmented-polarized transition.
}
\end{figure}

As $d$ is even further decreased, $\langle S_{max}\rangle/N$ vanishes
for $d < d_2$ in the thermodynamic limit, as shown in the inset of
Fig. \ref{fig:FSS_static}. The final state of the system is then {\em
fragmented} with a number of clusters saturating at $N$ as $d \to 0$
(see Fig. \ref{fig:FSS_no_clust_static}).  This polarized-fragmented
transition stems from the finite connectivity of the agents and is
akin to a percolation transition. It is indeed due to the fact that,
if the tolerance is too small, the probability for an agent to find
another neighboring agent with whom to interact vanishes; the
communication paths thus disappear from the network.  Let us consider
an agent $i$ with $k$ connections. The probability that a given
neighbor has initially an opinion within the tolerance range is simply
$2d$ so that the average number of neighbors with whom he can
communicate is $2dk$. The condition for the existence of percolating
paths of close enough opinions is thus simply $\bar{k} >
1/(2d)$, and the transition to fragmentation is expected at 
$d_1\approx 1/(2\bar{k})$.
Figure \ref{fig:diff_ks_static} displays the size of the
largest cluster in the final state as a function of $d$ for various
values of the average degree $\bar{k}$ of the network, showing indeed
that the polarized-fragmented transition occurs at a tolerance value
which scales as $1/\bar{k}$.

%%%%%%%%%%%%%%%%%%%%%%%%%%%%
%%%%%%%%%%%%%%%%%%%%%%%%%%%

We finally note that the polarized-fragmented transition is expected to
disappear if the interaction network is scale-free
with a diverging second moment of the degree
distribution \cite{pastorsatorras2004}. The
percolation transition indeed occurs at a vanishing threshold in such
networks in the thermodynamic limit. We have indeed checked (not shown) that
the polarized-fragmented transition is then shifted to much smaller
values of $d$, vanishing in the thermodynamic limit.

\section{Consensus formation on adaptive networks}
\label{sec:AN}

\begin{figure}[thb]
\vspace*{0.5cm}
\centering
\includegraphics[width=.45\textwidth]{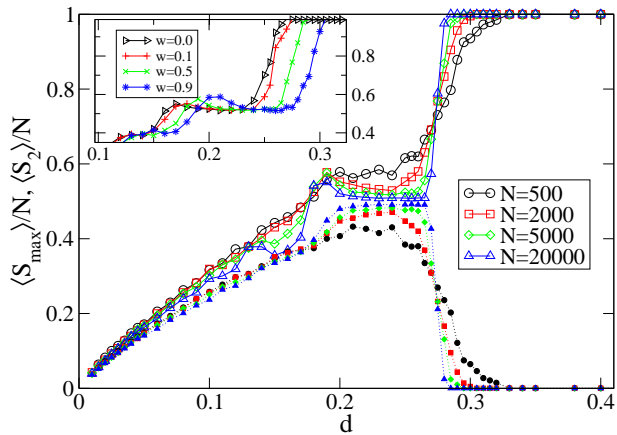}
\vspace*{-0.0cm}
\caption{ 
\label{fig:FSS_adaptive}(Color online)
Size of the largest (open symbols) and second-largest (filled symbols)
clusters as a function of the tolerance value on adaptive networks for
different system sizes ($\bar{k}=10$, $w=0.5$). The color coding is the
same for the first and second-largest cluster with respect to the system
size. Inset: maximal cluster size as a function of $d$ for different
rewiring rates.  The consensus-to-polarized transition point is shifted to
larger and larger tolerance values as the adaptivity of the networks
increases ($N=5000$, $\bar{k}=10$).}
\end{figure}

\begin{figure}[thb]
\vspace*{0.5cm}
\centering
\includegraphics[width=.45\textwidth]{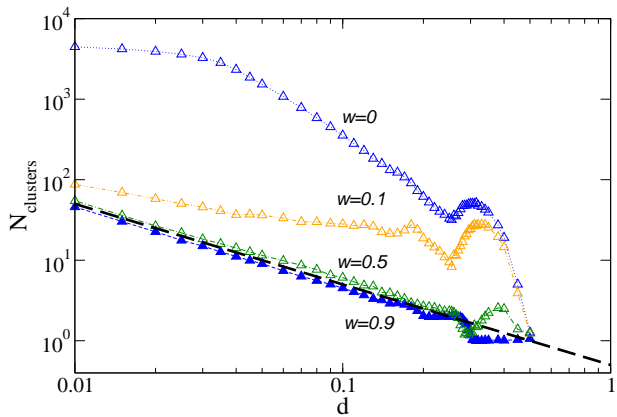}
\vspace*{-0.0cm}
\caption{ 
\label{fig:FSS_no_clust_adaptive}(Color online)
Number of clusters in the final state as a function of the tolerance
of the agents for different rewiring rates on adaptive networks with
average degree $\bar{k}=10$ and $N=5000$. The case of a static system 
($w=0$)
is also shown for reference. The dashed line corresponds to the value
$1/(2d)$, asymptotically valid at small tolerances in the mean-field case.
}
\end{figure}

Let us now turn to the case of adaptive network in which agents with far
apart opinions can break their connection. The rate of attempts to rewire
connections is given by $w$: the larger $w$ and the faster rewiring can
occur. Figure \ref{fig:FSS_adaptive} displays the sizes of the largest and
second largest clusters in the final state of the system, and
Fig. \ref{fig:FSS_no_clust_adaptive} shows the total number of clusters.

%%%%%%%%%%%%%%%%%%%%%%%%%%%%
%%%%%%%%%%%%%%%%%%%%%%%%%%%

At large enough tolerance, a unique cluster gathering all agents is obtained,
as in the static case. As the tolerance decreases, a consensus-to-polarized
transition is also observed, with the emergence of a second-largest cluster
with extensive size at $d_1(w)$. The jump in $\langle S_{max}\rangle/N$
becomes sharper as $N$ increases, indicating a first order transition as in
the static case.  Interestingly, we observe that $d_1(w)$ increases with $w$
(see inset of Fig. \ref{fig:FSS_adaptive}): the more easily agents can change
their connections, the larger tolerance values are necessary to achieve
consensus: agents can more easily search for other agents with whom they can
communicate, and break ties with the ones with too different opinions, so that
the formation of different clusters is favored.

%%%%%%%%%%%%%%%%%%%%%%%%%%%%
%%%%%%%%%%%%%%%%%%%%%%%%%%%

As $d$ decreases below $d_1$, a polarized phase is observed.  While
the sizes of the largest and second-largest clusters are close to the
case of a static network, important differences have to be
noted. First of all, in the adaptive network case, each opinion
cluster corresponds to a distinct connected component in the final
configuration. The network is therefore broken into $N_{clusters}$
disconnected components, while the static network remained connected
by definition. Moreover, the number of clusters is much smaller for
adaptive than for static networks, and decreases as $w$ increases, as
shown in Fig. \ref{fig:FSS_no_clust_adaptive}.

\begin{figure}[thb]
\vspace*{0.5cm}
\centering
\rotatebox{0}{
\includegraphics[width=.45\textwidth]{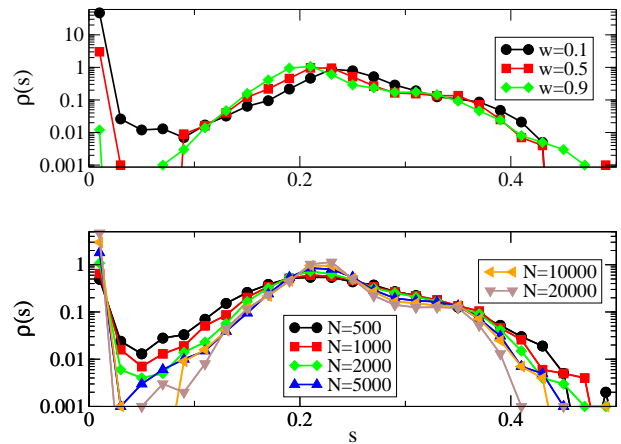}
}
\vspace*{-0.0cm}
\caption{ 
\label{fig:Pcs}(Color online)
Cluster size distribution
$\rho(s)$ for $d=0.1$ and $\bar{k}=10$, 
for various $w$ at $N=10^4$ (top) and for various sizes at $w=0.5$ (bottom).
}
\end{figure}

\begin{figure}[thb]
\vspace*{0.5cm}
\centering
\includegraphics[width=.45\textwidth]{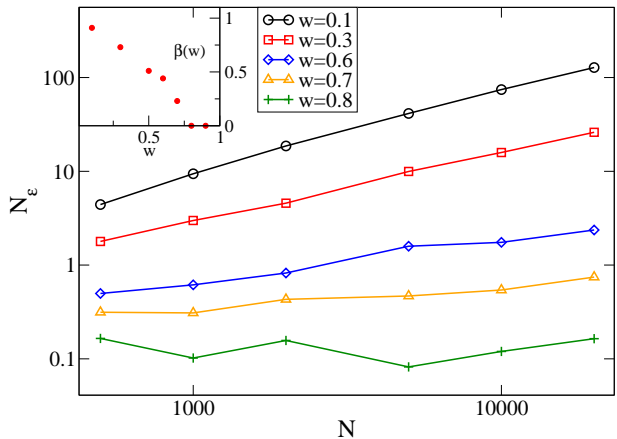}
\vspace*{-0.0cm}
\caption{ 
\label{fig:debris}(Color online)
Number of clusters with size less than $\epsilon=N/50$ for different
rewiring rates in systems with $\bar{k}=10$ and $d=0.1$. Inset:
measured exponents of the system-size dependent divergence of the number of
clusters ($\propto N^{\beta(w)}$) in the ``debris''. }
\end{figure}

More insight is given by the investigation of the normalized cluster size
distribution, shown in Fig. \ref{fig:Pcs} for $d=0.1$ and
$\bar{k}=10$. Similarly to the static case, it is formed of two parts,
\begin{equation}
\rho^w_N(s) \approx \dd(s)f^w(N)+Q^w(N,s),
\end{equation}
with $f^w(N)\propto N^{\beta(w)}$ as shown in Fig. \ref{fig:debris},
and $Q^w(N,s)$ is the distribution of clusters of macroscopic size
that converges to a regular finite distribution in
the large $N$ limit. 

%%%%%%%%%%%%%%%%%%%%%%%%%%%%
%%%%%%%%%%%%%%%%%%%%%%%%%%%

The first part of the distribution corresponds to {\em debris} of
finite-size. As can be seen in Fig. \ref{fig:debris}, on adaptive
networks ($w\ne0$) the expected size of the clusters in the debris is
increasing sublinearly with the system size, so that their weight is
vanishing compared to the system size, i.e. $f^w(N)/N \to 0$ in the
thermodynamic limit, while clusters of finite size compose a finite
fraction of the system in the static case ($\beta(w=0)=1$).

The polarized phase on adaptive networks is therefore different from
the one on static networks: thanks to the possibility of link rewiring, 
agents who would remain isolated (or in very small groups) 
on a static network may manage to find
agents with whom to communicate and thus enter a macroscopic cluster. 
Without rewiring on the other hand, a macroscopic number of agents remain
in fragmented components which coexist with few macroscopic clusters.

As shown in Fig. \ref{fig:Pcs}, the regular part of the cluster size
distribution shifts to smaller and smaller cluster sizes as the
rewiring rate $w$ increases. This phenomenon is similar to the 
shifting of the transition point between the
consensus and the polarized states: increasing the rewiring rate allows
agents to find more easily other agents with whom to communicate
and the formation of smaller clusters is favored.

%%%%%%%%%%%%%%%%%%%%%%%%%%%%
%%%%%%%%%%%%%%%%%%%%%%%%%%%

%%%%%%%%%%%%%%%%%%%%%%%%%%%
\begin{figure}[thb]
\vspace*{0.5cm}
\centering
\includegraphics[width=.45\textwidth]{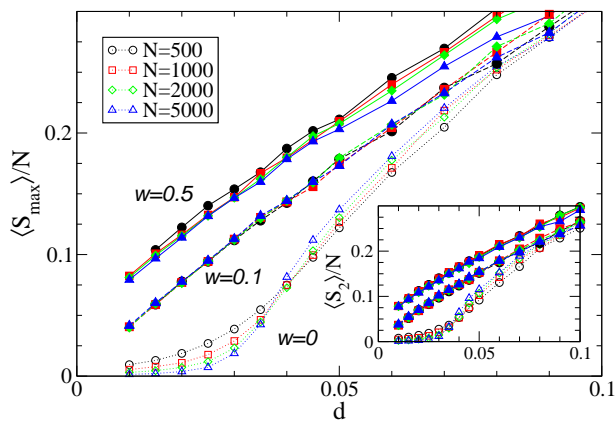}
\vspace*{-0.0cm}
\caption{ 
\label{fig:FSS_adaptive_bis}(Color online)
Size of the largest cluster as a function of the tolerance value on
adaptive networks (filled symbols) for different system sizes
($\bar{k}=10$), compared with the same quantity on static network (open
symbols). The curves for $w=0.5$, otherwise collapsing with the $w=0.1$
ones, are shifted for clarity. Inset: same for the second-largest
cluster.}
\end{figure}
%%%%%%%%%%%%%%%%%%%%%%%%%%%

Figure \ref{fig:FSS_adaptive_bis} finally compares adaptive and static
networks for small values of the tolerance. Strikingly, the fragmented
phase disappears as soon as the rewiring of the links is enabled. The size
of the largest component decays smoothly as the tolerance decreases, but
remains extensive, in contrast with the static case. Rewiring processes
thus allow the small clusters to group together and reach extensive sizes
even below the polarized-to-fragmented transition, present only on static
networks.

The comparison of the static and adaptive cases shows that the ability
of the network to adapt as a consequence of the opinion dynamics 
has strong consequences, which interestingly are somehow opposite
in the various tolerance ranges and for the various types of clusters. 
On the one hand, large clusters can be more easily broken by rewiring,
and global consensus is more difficult to reach. At intermediate
tolerance values, the extensive clusters are smaller when rewiring
is enhanced. On the contrary, the small clusters of finite size 
have opportunities to find agents with whom to communicate, and 
therefore to merge with large clusters, leading to a strongly decreased
total number of clusters and to a true polarized state instead of a mixture
of polarization and fragmentation. 
At very small tolerance values finally, the fragmentation transition even 
disappears, and extensive clusters are obtained for arbitrarily small
tolerance.

\section{Temporal behavior}
\label{sec:Temp}

In the previous sections, we have focused on the final state
in which the system settles as a result of the interplay between
opinions' and network's dynamics. In order to better understand
the role of the different processes, we now consider the time evolution
of the clusters of agents. For the static network, only opinion clusters
evolve. On the other hand, topological and opinion clusters are not
identical for adaptive network: a given connected component 
of the network can host several opinion clusters.

First, as a reference, let us consider the temporal behavior on static
networks: In this case, the time of convergence depends on the distance of
$d$ from the fragmented-to-polarized transition, $d_2$. For large tolerance
values, $d_2<d$, the convergence time grows linearly with the system size
(data not shown) and it increases as $d \to d_2$ since, in this limit in
the percolating opinion-clusters, the average length of the communication
path between two nodes increases \cite{Lopez2007} and the clusters are
becoming more and more tree-like. Near $d_2$, the average length of the
communication channel between two nodes in an opinion cluster grows as a
nontrivial power of the system size \cite{Braunstein2003} which in turn
raises the possibility of a convergence time which grows faster than
linearly with the system size (not shown). For $d<d_2$, the convergence
time grows linearly with logarithmic corrections with respect to the system
size. Though, it decreases as the tolerance of the agents decreases since
the size of the tree-like opinion clusters also decreases. For very small
tolerance values ($d \ll d_2$), the system in fact almost does not evolve
since agents rarely find neighbors with whom they can communicate.

\begin{figure}[t]
\vspace*{0.5cm}
\centering
\rotatebox{0}{
\includegraphics[width=.5\textwidth]{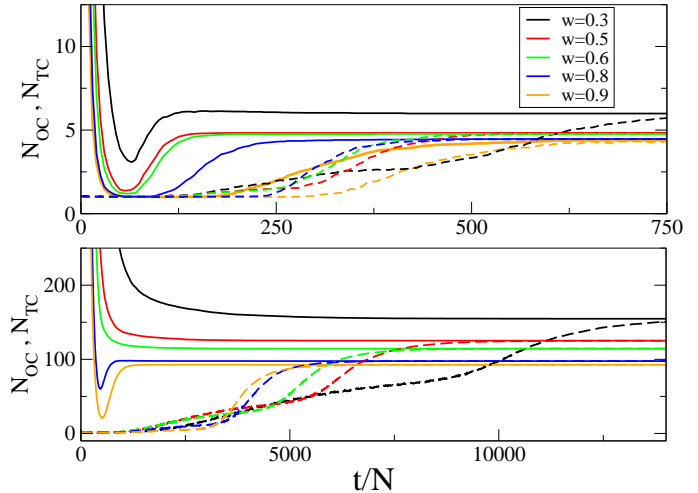}
}
\vspace*{-0.0cm}
\caption{ 
\label{fig:temp_1}(Color online) 
Temporal behavior of the system with high (top: $d=0.1$, $\bar{k}=10$, and
$N=10^3$), and low (bottom: $d=0.005$, $\bar{k}=10$, and $N=10^4$)
tolerance of the agents. (full lines: number of opinion clusters; dashed
lines: number of topological clusters) }
\end{figure}

\begin{figure}[t]
\vspace*{0.5cm}
\centering
\includegraphics[width=.45\textwidth]{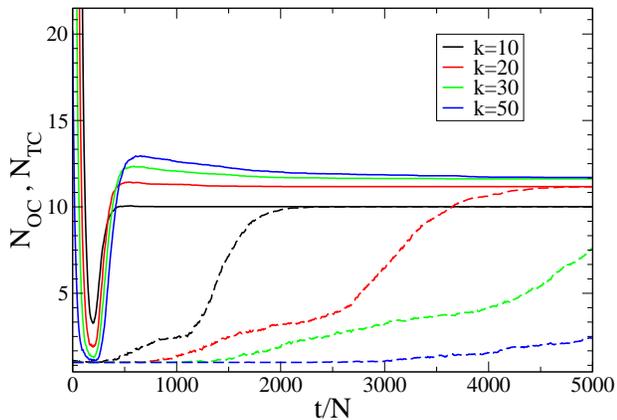}
\vspace*{-0.0cm}
\caption{ 
\label{fig:temp_3}(Color online) 
Temporal behavior of the system for fixed $w$ and $d$ and varying $\bar{k}$
($N=2000$, $w=0.5$, and $d=0.05$). For these parameter values, the
estimation in the text predicts larger and larger separation between 
$t_l$ and $t_o$, which is confirmed by the simulation data. (full lines:
number of opinion clusters; dashed lines: number of topological clusters) }
\end{figure}

In the case of adaptive networks, Fig.~\ref{fig:temp_1} shows the evolution
of the number of opinion clusters $N_{OC}$ and of topological clusters
(connected components) $N_{TC}$. The figures clearly show the existence of
three different timescales for the clusters' evolution.  In the initial
configuration, a large number of separate opinion clusters are found,
corresponding to percolation clusters; their number is naturally larger for
smaller tolerance values and smaller $\bar{k}$. The early-time evolution is
then mostly determined by the adaptive nature of the network which allows
agents to look for other agents with similar opinions. The importance of
this early-time behavior can be particularly emphasized in the small
tolerance regime which would lead to a fragmented state for a static
network. As shown in Fig. \ref{fig:temp_1} indeed, the number of opinion
clusters decreases very fast, from an extensive to a finite number, while
the network is still globally connected ($N_{TC}=1$). The possibility of
rewiring connections allows therefore to form percolating/macroscopic
clusters even at small tolerance values, which explains the disappearance
of the fragmented phase.  The characteristic time of this phenomenon is
given by the time necessary for an agent to find at least one partner ``to
be able to communicate with,'' $t_f$.  If the rewiring rate $w$ is large,
$N_{OC}$ reaches a minimum before increasing again when opinions evolve. An
opinion cluster indeed hosts agents that are connected by a path of
potential communication, but even neighboring agents' opinions can evolve
and drift further apart due to interactions with other neighbors. An
opinion cluster can therefore divide itself into several clusters because
of the opinion dynamics, and $N_{OC}$ increases. The corresponding
timescale $t_o$ describes the formation of groups of agents with identical
opinions on a still connected network.  Finally, after the formation of
groups with uniform opinions, the number of topological clusters increases
and converges to the number of opinion clusters. This last phase therefore
corresponds to a breaking of the links which join opinion clusters with
different opinions. The timescale of this final change, $t_l$,
characterizes the time needed by an agent to rewire its links with agents
out of his tolerance range towards agents with the same opinion. Depending
on the parameters of the system, this final regime can take place at
timescales either much larger than those associated with the opinion
cluster formation, or on similar timescales. The two possibilities are
illustrated in both Fig.s \ref{fig:temp_1} and \ref{fig:temp_3}.
The case of widely separated timescales allows to consider
that the opinion dynamics and rewiring process occur independently, with
links evolving between fixed opinion clusters; further investigation is
then possible and a mean-field analysis allows to gain insight into the
clusters' topological structure \cite{AGandBK}.

%%%%%%%%%%%%%%%%%%%%%%%%%%%%
%%%%%%%%%%%%%%%%%%%%%%%%%%%

The dependence of the various timescales on the parameters can be estimated
as follows: 

i) Initially, there is always a finite fraction of separated nodes, loners,
which are not members of any percolating opinion clusters both below and
above the fragmented-to-polarized transition. Therefore the time
characterizing the transition of $N_{OC}$ from an extensive to a finite
value is $t_f$. A typical loner has no neighbor with whom to
communicate in the initial configuration. The time needed to rewire any of
its links is proportional to $1/w$, and the probability to find an agent
within tolerance range is $2d$, so that
\begin{equation} 
t_f \propto 1/(w d). 
\end{equation}

ii) The estimation of the timescale of the opinion evolution is a more
complex task. In all cases, opinions evolve at a rate $(1-w)$, so that $t_o
\propto \tau_o/(1-w)$, where $\tau_o$ is yet to be characterized. For low
rewiring rates and when $d \ll d_2$, the success rate of the discussions
are determined by the ratio of newly found friendly neighbours and the
average degree of the node. Therefore, the larger is this ratio, the
smaller is $\tau_o$.  When $d \gg d_2$, a typical agent can successfully
communicate with the $2d$ fraction of his friends irrespectively of his
degree. Therefore $\tau_o$ can only depend weakly on $\bar{k}$ in this case
(see Fig. \ref{fig:temp_3}).  Before considering the behavior of $t_o$ for
high rewiring rates, the scaling of $t_l$ can be estimated as follows:

iii) For any $d$ values, links are updated with frequency $w$. Let us
consider a typical opinion cluster.  The number of its links which need to
be rewired is proportional to the total number of links ($\propto \bar{k}
N$), and to the amount of opinions outside of the tolerance range ($\propto
(1-2d)$). The probability to rewire towards a close enough opinion is
moreover $\propto d$ so that
\begin{equation}
t_l \propto \bar{k}(1-2d)/(w d) .
\label{tl}
\end{equation}
For low rewiring rates, $t_l$ is the longest timescale of the time evolution
therefore the convergence time also scales as $t_l$.

For high rewiring rates, it is possible for a typical agent to successfully
rewire all his links to point to agents with tolerable opinions before
committing himself to changing his own opinion. This situation takes place
when $t_l$ is less than or comparable to $1/(1-w)$. In this case, almost
all of the negotiations are successful and both the convergence time and
$t_o$ are expected to be $\propto 1/(1-w)$. Though, observations show that
the situation is more convoluted: as the tolerance of the agents increases,
fewer and fewer opinion clusters are present in the system. Nevertheless,
the disappearance of a cluster in many cases happen by an initial
unsuccessful attempt to form two or more separate clusters which eventually
merge into one. During this initial evolution most of the links are broken
between these communities and the convergence to a common opinion is only
mediated by few individuals connected to both groups (see for example the
red curve in Fig. \ref{fig:path_adaptive}). These individuals form a narrow
channel of communication between the two communities throughout the process
resulting in very long convergence times in certain regimes of
tolerance values.

Similar merging of communities can be also be observed on static networks
though their behavior is less drastic (see Fig. \ref{fig:path_static}).
Even if the merging is started by few individuals, similarly to what
happens on adaptive networks, as soon as the average opinion of the two
communities become close to each other, the members of the two communities
suddenly engage in fruitful discussions and their opinions converge rapidly
to a uniform value.

Finally, Fig. \ref{fig:temp_1} illustrates how the timescale of opinion
evolution increases when $w$ increases and how the separation between $t_o$
and $t_l$ increases as the tolerance $d$ is reduced, so that a change of
parameters can lead from similar to well separated timescales.  Figure
\ref{fig:temp_3} moreover shows that an increase in the average degree also
leads to more and more separated timescales, as seen from 
the arguments in paragraph ii) and Eq. (\ref{tl}).

%%%%%%%%%%%%%%%%%%%%%%%%%%%%
%%%%%%%%%%%%%%%%%%%%%%%%%%%

\section{A variant of the model}
\label{sec:variant}

The Deffuant model considers agents that have a certain tolerance range and
can strictly not communicate with agents having opinions outside this range.
This drastic behavior can seem unrealistic, and we consider here a variation
of the model in which agents still have a finite probability to communicate
event if their initial opinions are far apart. While various extensions of the
update rules could be considered, we limit our study to the following simple
generalization of the model: if agents $i$ and $j$ have close enough opinions,
i.e., if $|o(i)-o(j)|<d$, they adopt the same intermediate opinion
$\frac{o(i)+o(j)}{2}$; if $|o(i)-(o(j)|>d$ on the other hand, 
the opinions of the two agents
converge to their mean value with probability 
\begin{equation}
\label{eq:probabilistic}
 p=e^{1-\frac{|o(i)-o(j)|}{d}} \;.
\end{equation}
In the exponent, the $1$ term is present to make the probability, $p$,
continuous when $|o(i)-o(j)|=d$. The rewiring rule is also changed: two
agents may break their connection to each other only if they are outside of
each others tolerance range and, in this case, they do it with probability
\begin{equation}
 p_{rw}=1-e^{1-\frac{|o(i)-o(j)|}{d}} \;.
\end{equation}

\begin{figure}[t]
\vspace*{1.0cm}
\centering
\includegraphics[width=.45\textwidth]{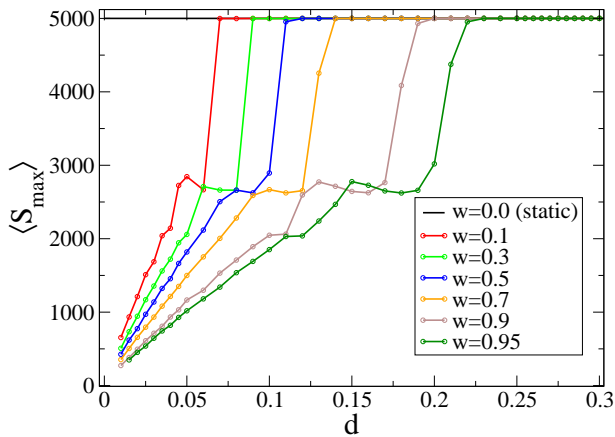}
\vspace*{-0.0cm}
\caption{ 
\label{fig:exp_max_clust}(Color online)
Maximal cluster size as a function of the tolerance of the agents for a
variant of the model in which even agents with opinions outside of
each other tolerance range can communicate with a probability given by
Eq. (\ref{eq:probabilistic}).  }
\end{figure}

In this probabilistic model, consensus is always achieved on any static
network, since any couple of neighboring agents always have some probability
to communicate and reach the same opinion. Decreasing the tolerance of the
agents only increases the corresponding convergence time (not shown). On
adaptive networks however, the rewiring rule allows opinion clusters to
separate, and a picture similar to the one of the original Deffuant model is
obtained, as shown in Fig.  \ref{fig:exp_max_clust}, with a transition between
a consensus state at large tolerance to a polarized state as $d$ decreases.
The transition is also shifted to larger and larger tolerance values as $w$
increases. It is interesting to compare the effect of rewiring on the system
with the original consensus-formation rules and in the case of this variant
when $d<d_2$: in the original model, rewiring drove the system to a {\em more
  homogeneous} (polarized) state than that observed on static networks (where
fragmentation is obtained); while, in this variant of the model, rewiring
drives the system to a {\em more inhomogeneous} state (polarized) than that on
static networks (which is a consensus reached in very long times).

Interestingly, the behavior of the model on adaptive networks is in fact
more robust than on static networks, since the same global picture
is observed for strict or probabilistic communication rules, while
a strong difference is obtained on static networks.
 
%%%%%%%%%%%%%%%%%%%%%%%%%%%%
%%%%%%%%%%%%%%%%%%%%%%%%%%%

\section{Conclusions}

In this paper, we have studied consensus formation on static and adaptive
networks through the investigation of a simple model of opinion dynamics with
bounded confidence: agents with close enough opinions reach an agreement,
while they can not communicate if their opinions are too far apart.  When the
agents are linked through a static interaction network, two transitions are
found: at large tolerance values, a global consensus is reached; intermediate
tolerance leads to a coexistence of several extensive groups or clusters of
agents sharing a common opinion with a large number of small (finite-size)
clusters. At very small tolerance values finally, a fragmented state
is obtained, with an extensive number of small groups. 
This is in contrast with the mean-field case in which the number of
groups is, roughly speaking, the inverse of the tolerance range. 

When agents can rewire their links in a way depending on the opinions
of their neighbors, i.e. break connections with neighbors with far apart
opinions, the situation changes in various ways. At large tolerance
values, the polarization transition is shifted since rewiring makes
it easier for a large connected cluster to be broken in various
parts. The possibility of network topological change therefore
renders global consensus more difficult to achieve. On the other hand,
for smaller tolerance values, the number of finite-size clusters is
drastically reduced since agents can more easily find other agents
with whom to reach an agreement. A real polarized phase is thus obtained,
and the transition to a fragmented state is even suppressed: extensive
clusters are obtained even at very low tolerance. 

The detailed investigation of the system's time evolution reveals that the
rewiring dynamics plays an important role both at early and late times: at
early-times, adaptive rewiring enhances communication between agents and
fosters giant cluster formation while, at late times, adaptation results in
the breakup of the network into separate clusters after (or while) opinions
evolve locally. The various involved timescales depend on the model's
parameters and can be either well separated or similar.

Finally, we have considered changes in the microscopic rule of opinion
evolution, from a strict and maybe unrealistic rule of sharp
tolerance threshold to a smoother decrease of communication when
opinions are further apart. Interestingly, such a change has a dramatic
effect when the interaction network is fixed, since the system
then always reaches consensus. The scenario of adaptive networks is however
more robust, with a transition between consensus and polarized states
as the tolerance is decreased. This emphasizes the relevance of
considering the possibility of evolving topologies when studying the emergence
of collective behavior in models for opinion formation.

Further investigations will consider the detailed topological structure of
the clusters or groups of agents sharing the same opinion \cite{AGandBK},
and the evolution of other models for opinion dynamics on adaptive
networks, where for example bounded confidence is either absent (such as
the Voter model) or replaced by negotiation processes (such as the Naming
Game). In such models, the asymmetry of the relation between the agents
involved in an interaction has been shown to have further interesting
effects on static networks \cite{Suchecki:2005a,castellano,Dallasta:2006c},
and can be expected to couple with the network evolution with new
relevant consequences \cite{inprep}.

\end{document}